\documentclass[conference]{IEEEtran}
\IEEEoverridecommandlockouts
\usepackage{graphicx}
\usepackage{color}
\usepackage{colortbl}
\usepackage{todonotes}
\usepackage{multirow}
\usepackage{booktabs}
\usepackage{longtable}
\usepackage{comment}
\usepackage{tikz}
\usetikzlibrary{shapes,arrows,positioning}
\usepackage{float}
\tikzset{decision/.style={diamond, draw, fill=yellow!20, 
    text width=6em, text badly centered, node distance=3cm, inner sep=0pt},
block/.style={rectangle, draw, fill=gray!20, 
    text width=12em, text centered, rounded corners, minimum height=4em},
small_block/.style={circle, draw, fill=white!20, 
    text width=0em, text centered, rounded corners, minimum height=0em},    
line/.style={draw, -latex'},
cloud/.style={draw, ellipse,fill=red!20, node distance=3cm,
    minimum height=2em},}
\usepackage{tikz-dependency}
\def\checkmark{\tikz\fill[scale=0.4](0,.35) -- (.25,0) -- (1,.7) -- (.25,.15) -- cycle;}

\usepackage{xr}
\makeatletter
\newcommand*{\addFileDependency}[1]{
  \typeout{(#1)}
  \@addtofilelist{#1}
  \IfFileExists{#1}{}{\typeout{No file #1.}}
}
\makeatother

\newcommand*{\myexternaldocument}[1]{%
    \externaldocument{#1}%
    \addFileDependency{#1.tex}%
    \addFileDependency{#1.aux}%
}
\myexternaldocument{supplemental-material}
\begin{document}
\title{Evaluating the Role of Security Assurance Cases in Agile Medical Device Development}

\author{
\IEEEauthorblockN{Max Fransson\IEEEauthorrefmark{1}, Adam Andersson\IEEEauthorrefmark{3}, Mazen Mohamad\IEEEauthorrefmark{1}\IEEEauthorrefmark{3}, Jan-Philipp Steghöfer\IEEEauthorrefmark{2}}
\IEEEauthorblockA{\IEEEauthorrefmark{1}Research Institutes of Sweden (RISE), Gothenburg, Sweden. \emph{max.fransson@ri.se}, \emph{mazen.mohamad@ri.se}}
\IEEEauthorblockA{\IEEEauthorrefmark{2}XITASO GmbH IT \& Software Solutions, Augsburg, Germany. \emph{jan-philipp.steghoefer@xitaso.com}}
\IEEEauthorblockA{\IEEEauthorrefmark{3}Chalmers University of Technology, Gothenburg, Sweden. \emph{adande@student.chalmers.se}}
}

\maketitle
\begin{abstract}
Cybersecurity issues in medical devices threaten patient safety and can cause harm if exploited. Standards and regulations therefore require vendors of such devices to provide an assessment of the cybersecurity risks as well as a description of their mitigation.
Security assurance cases (SACs) capture these elements as a structured argument. Compiling an SAC requires taking domain-specific regulations and requirements as well as the way of working into account.
%
In this case study, we evaluate CASCADE, an approach for building SAC in the context of a large medical device manufacturer with an established agile development workflow. We investigate the regulatory context as well as the adaptations needed in the development process.
Our results show the suitability of SACs in the medical device industry. We identified 17 use cases in which an SAC supports internal and external needs. The connection to safety assurance can be achieved by incorporating information from the risk assessment matrix into the SAC. Integration into the development process can be achieved by introducing a new role and rules for the design review and the release to production as well as additional criteria for the definition of done. We also show that SACs built with CASCADE fulfill the requirements of relevant standards in the medical domain such as ISO~14971.
\end{abstract}


\section{Introduction}
\label{sec:introduction}



Software is increasingly responsible for therapeutic choices~\cite{ema_medical_2018} and the devices it is running on become more connected~\cite{medical_iot}. This development introduces cybersecurity risks which have the potential to impact the health and safety of patients~\cite{iso_62304}.
Cybersecurity and safety in the medical domain are therefore regulated by bodies such as the European Medicines Agency (EMA) and the Food and Drug Administration (FDA). Regulations include those aimed at protecting sensitive assets such as personal medical data (HIPAA) as well as the potentially hazardous impact on the patient using the device~\cite{iso_14971}. Risk management is particularly important for getting market approval for medical devices~\cite{premarket-fda,postmarket-fda}. 

To comply with relevant regulations, a vendor needs to argue that their system is secure. A common technique is to break the system into its constituent parts and then build a claim about the security of the entire system out of the claims about the security of the parts. This kind of bottom-up argument is one of the primary elements of a Security Assurance Case (SAC)~\cite{weinstock_arguing_nodate}.
SAC are an emerging approach for assuring the security of software systems in different domains~\cite{mohamad_security_2021}.
They collect all claims about the security of a system and its parts, as well as about the capabilities and the maturity of an organisation along with the evidence that supports these claims. This self-assessment of both the product as well as the organisation is often recorded using Goal Structured Notation (GSN)~\cite{weinstock_arguing_nodate}. The claims are structured to form an argument whose validity is supported by evidence.



Once an SAC is created, it needs to be maintained to keep up with the changes made to the system. The security of the system's constituent parts and of its overall security needs to be re-evaluated based on claims and evidence. This requires resources, attention and relevant knowledge from the system's maintainers and often occurs in the context of an agile development approach. 
While there are agile methods that include security activities~\cite{rindell2017busting,rindell2021security}, they do not yet support formalising the self-assessment of the product and the organisation and providing a cohesive, evidence-based argument for the security of the system under construction. 
Creating and maintaining an SAC closes this gap, but needs to be integrated into the development approach in a suitable form that allows developers to continue delivering value regularly. 

There are multiple studies concerning the creation and maintenance of SAC in the medical domain \cite{medical4,medical5,medical6}, however, there is a need for further validation in industrial contexts \cite{mohamad_security_2021}.
In this paper, we report on our experiences with the creation of SAC in the agile development of a medical device using the CASCADE~\cite{mohamad2023cascade} approach for building SACs. We focused on the applicability of SAC in the medical domain and if there were any domain-specific adaptations necessary. We also evaluated how SACs integrate into an agile development process where features and components are iteratively added and changed. This lead us to two research questions:

\begin{description}
    \item[RQ1:] How applicable are SAC in the development of medical devices?
    \item[RQ2:] How can the creation and maintenance of SAC be integrated into an agile development process?
\end{description}

To answer the RQ, we conducted a case study with a major international vendor of medical devices. We were embedded in a department creating a platform (HOOP) that manages patient and medical practitioner data in the context of clinical studies. We specifically looked at an app that is used to collect sensitive medical information in ongoing clinical trials and is classified as a medical device. We developed an SAC for the app and evaluated our work with interviews and focus groups.


\section{Background and Related Work}
\label{sec:background}



\subsection{Security Assurance Cases}
\label{sec:background:sac}

A Security Assurance Case (SAC) is a structured body consisting primarily of security claims and evidence for these claims~\cite{weinstock_arguing_nodate}. 
A top claim for the security of the entire system is refined to security sub-claims. This is a multi-stage process that can use strategies as intermediary steps to describe on what basis decomposition is performed. This is repeated until the claims are broken down to a sufficiently granular level, where concrete evidence can be assigned to them. When evidence has been assigned for all low-level claims, and given that the derived claims are exhaustive for the system, the case contains the argument for the existing security posture of the system. An excerpt of an SAC can be found in Figure 1 in the supplemental material~\cite{supplemental-material}.





\begin{figure}[tb]
\centering
\includegraphics[width=0.9\columnwidth]{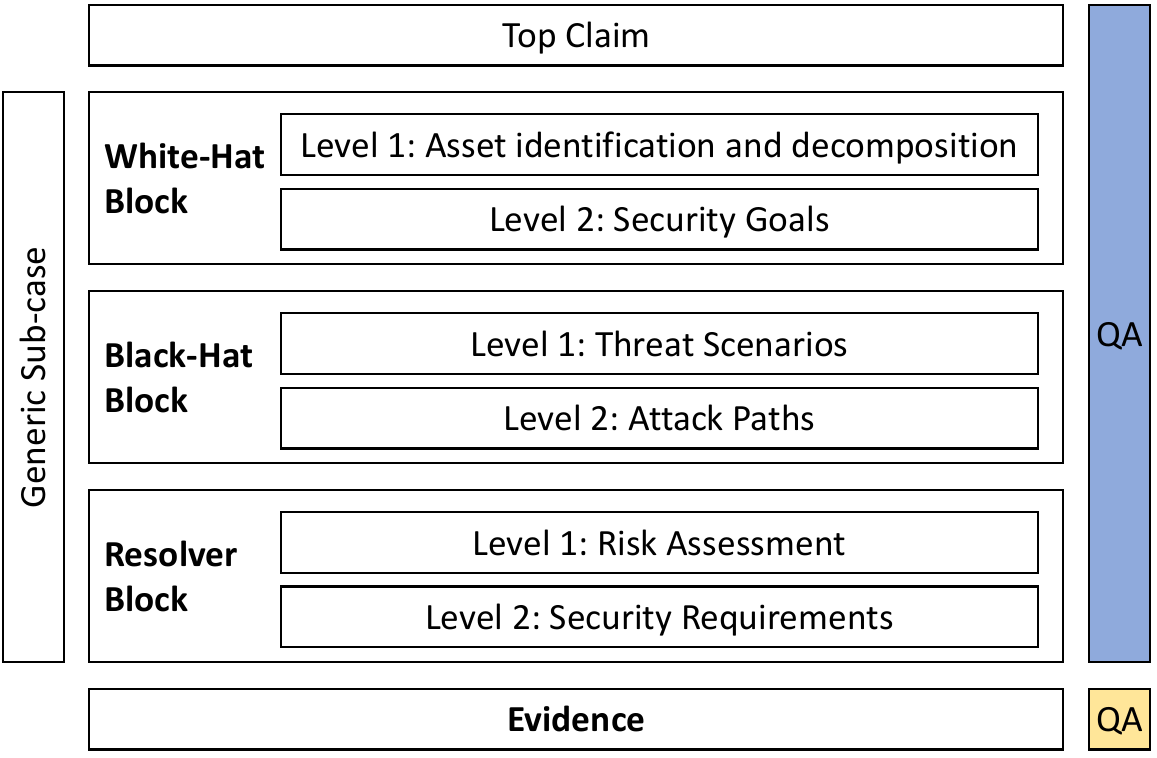}
\caption{The structure for the CASCADE approach.}
\label{fig:cascade_blocks}
\vspace{-0.1cm}
\end{figure}

\textbf{CASCADE}
is an approach by Mohamad et al.~\cite{mohamad2023cascade} for creating security assurance cases based on the block structure depicted in Figure~\ref{fig:cascade_blocks}. CASCADE uses a top claim and evidence and divides the rest of the case into three blocks and a generic sub-case to provide structure and an inherent flow for the SAC.

\emph{White hat block:}
Contains all identified assets, i.e., important artefacts in the system and their decomposition. Security goals describe a security property, e.g., stem from the confidentiality, integrity, and availability (CIA) triad, that should be preserved for the decomposed asset. 


\emph{Black hat block:}
Defines claims about threat scenarios that could compromise the security goals defined in the white hat block. Also specifies claims about attack vectors that could realise these threat scenarios. 


\emph{Resolver block:}
Defines claims about the risk treatment and then provides specific security requirements that need to be fulfilled for the treatment to be effective. Valid treatments would be to \emph{accept}, \emph{mitigate} or \emph{transfer} the risk. 


\emph{Generic sub-case:}
Contains claims that are not specific to the case and could apply to other systems at the company as well. Some usages for this are to express things like ``mandatory security training'', ``company security policies'', ``thorough firewalls'' etc.


In addition to this block structure, CASCADE extends conventional SACs by incorporating quality claims to argue for the completeness of the decomposition of sub-claims as well as for the quality of the provided evidence.
%
%
In terms of completeness, the aim is to provide assurance that the strategy used has decomposed all relevant sub-claims and that it has taken the relevant aspects and assets into consideration. 
In terms of quality, the aim is to argue for the soundness and quality of all evidence assigned to a claim. 

\subsection{Standards and guidelines in the medical domain}
\label{sec:background:standards}

There are a number of international standards that govern how systems in the medical domain and in particular for diagnosis and treatment have to be developed and validated.

\emph{Software as a Medical Device} (SaMD)~\cite{imdrf_samd_2017} imposes requirements on different types of software used in a medical device throughout the life cycle of a product, i.e., both before and after the release of a product onto the market. A particular focus of SaMD is trustworthiness. 
%
%
The FDA publishes guidelines for the ``\emph{pre-market}'' and ``\emph{post-market}'' product development phases which need to be followed to be SaMD compliant~\cite{premarket-fda,postmarket-fda}. Both documents encourage manufacturers to address risks in different stages of the product life cycle to lower the likelihood that any of them will cause concerns to users' health when the device is in use.
Similarly, the ``Guidance on Cybersecurity for medical devices''~\cite{mdcg_guidance} issued by the Medical Device Coordination Group (MDCG, a group of experts convened by the EU) focuses on guiding manufacturers on how to 
ensure that new devices are built using state-of-the-art  risk management and cybersecurity.

ISO~14971~\cite{iso_14971} specifies requirements for the application of risk management to medical devices and is often used by regulatory authorities to issue market approval. A central piece of ISO~14971 is the ``risk management file'' which provides traceability for hazards and risks identified by a risk management process. 
The standard is supported by ISO 24791~\cite{iso_24791} which contains guidelines for addressing  hazardous situations (including those that could result in physical harm) that could stem from cybersecurity risks.
%
%

ISO~62304~\cite{iso_62304} also mentions the risk management file and makes a number of additions to the risk management process defined in ISO~14971~\cite{iso_14971}. 
It introduces ``Software Of Unknown Provenance'' (SoUP) and highlights SoUP as a potential cause for failure and unexpected results. SoUP are software artifacts in a medical device that are not developed by the company manufacturing the medical device and might not have been developed with the same rigour as the rest of the device's software.
%
There is an emphasis on traceability of the hazard (safety) risk and the risk control measure as well as evidence of the verification of the risk control measure.

\subsection{Related work}






Arnab Ray and Rance~\cite{ray_security_2015} present security assurance cases as an approach to increase security and therefore safety for medical devices. 
They propose that SACs can be a driving force for the design, implementation, verification, and documentation of medical devices. By creating the case alongside  development, desirable security practices can be achieved and the case itself may be used as a representation of and motivation for device security. This is in contrast to current observed usage where assurance cases for medical devices are often created after development as a means to satisfy regulatory requirements. We adopt these ideas when integrating security work in the existing process at the case company. 




Beznosov and Kruchten~\cite{beznosov2004towards} categorise security assurance practices in terms of whether they match with agile approaches and identify a number of ``mismatches''. However, they assume a ``pure'' agile approach such as Scrum or XP rather than one that already considers safety-critical systems as in the case company which already provides the rigour required for security assurance.

Scrum as an agile approach in the context of security is considered by 
Ben Othmane et al.~\cite{othmane_using_2014}. 
The authors highlight key concerns when using SAC in conjunction with an iterative approach: component updates could invalidate previously created SAC claims; new components require re-evaluation of all related SAC claims; changing the use context of the software requires re-evaluation of all related SAC; and adding a new claim requires re-evaluation of all related SAC claims.
They propose creating a high level architecture as well as an incremental road map to plan work on a security feature. This prevents the redevelopment of security mechanisms already in place and applying the knowledge gained in previous steps.
These findings are also supported by Oueslati et al.~\cite{oueslati2015literature} who performed a systematic literature review on the issue.





\subsection{Case Study environment}
The case study is performed at a large multinational pharmaceutical company in the medical domain with more than 50,000 employees worldwide. The company has increased its focus on the production of software, often in the context of data gathering for clinical studies or for medical devices. The department in which we conducted the study is building the \emph{HOOP} platform to handle patient and medical practitioner data for clinical studies. 

We focus our work on an app based on HOOP that facilitates data collection in ongoing clinical studies. This app generates a large amount of sensitive medical data about individual patients. The data is used by several services and components that have distinct purposes. The data that each component needs varies, and there is sometimes a need for duplication of data for these components to work correctly.
Since the app is highly integrated with other services and because it handles sensitive data, 
it is subject to the regulations in the medical device domain and conformance to these regulations and applicable laws must be shown by the developing organisation. Therefore, the company treats the platform HOOP itself and the app we focus on as a medical device.
In the following, we do not distinguish between the platform and the app when we mention HOOP.

There are a number of relevant roles involved in the development of HOOP, including an ``SaMD Quality Lead''. An overview of the roles relevant for this study is listed in Table 1 in the supplemental material~\cite{supplemental-material}.
A \emph{system design document}~\cite{iso_12207}, also called a ``blueprint'', is kept for HOOP. This document contains all information about the system architecture and design. The blueprint is incrementally updated during the product's life cycle as part of the iterative workflow at the company. This means that new features and artefacts introduced in the software system design phase have to be reflected in the blueprint. 





\section{Research Method}
\label{sec:methods}

We conducted a case study in the field with the goal of investigating the suitability of SACs in the medical domain, and the ability of the created case to be maintained in an agile workflow. This allowed us to use the real-world context and setting at the case company~\cite{stol_abc_2018}.

\begin{figure}[tb]
\centering
\includegraphics[width=0.95\columnwidth]{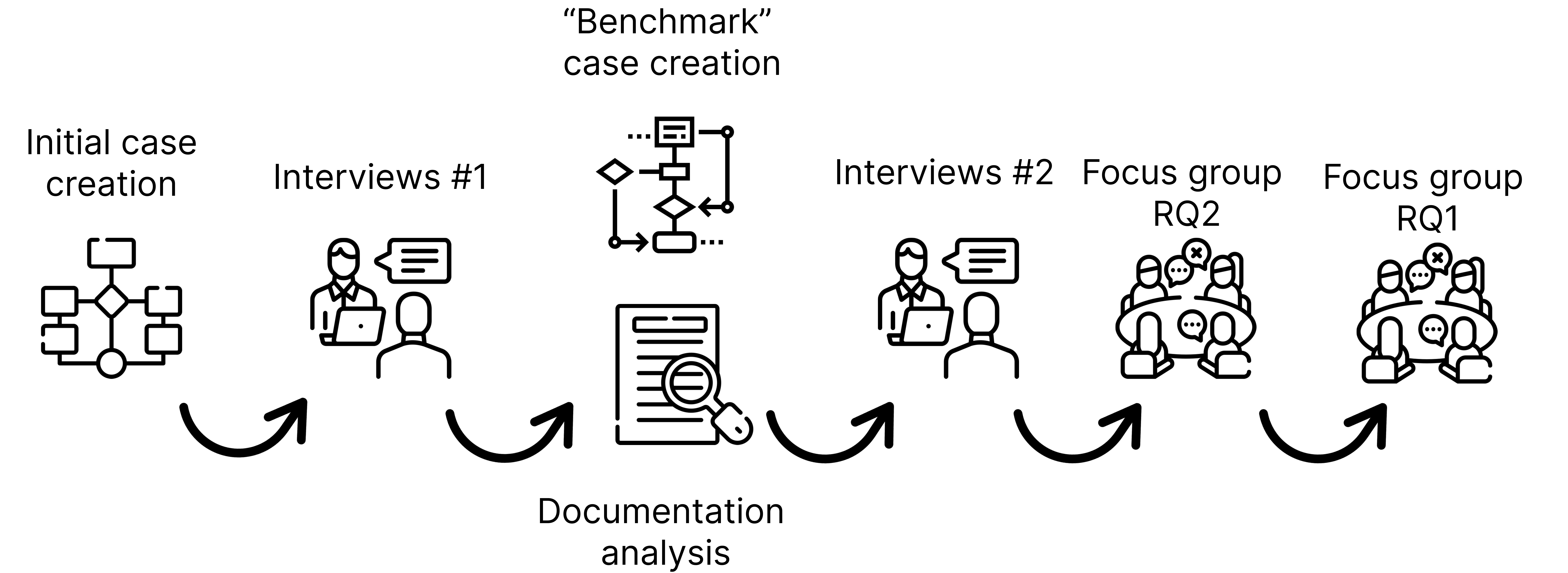}
\caption{The methods used for the case study in chronological order.}
\label{fig:method_figure_rq1}
\end{figure}

We conducted the activities outlined in Figure~\ref{fig:method_figure_rq1}.
We created an SAC 
for use in our first round of interviews to elicit the system information necessary to create the benchmark case as well as elicit the documents for the document analysis. The second round of interviews served as initial validation for the document analysis results and as data collection for RQ2. We then conducted two focus groups to confirm the results for the two research questions.

\noindent\emph{Participants.}
We included six different actors in the case study (cf.~Table~\ref{tbl:participants}). The participants roles  varied significantly, which helped to encourage discussions and provide opinions grounded in different parts of the studied system. 


\begin{table}[tb]
\caption{Participants in the case study activities}
\label{tbl:participants}
\resizebox{\columnwidth}{!}{
\begin{tabular}{@{}llllllllll@{}}
\toprule
            {} & {Role} & \rotatebox{90}{Interviews \#1} & \rotatebox{90}{Benchmark case} & \rotatebox{90}{Interviews \#2} & \rotatebox{90}{Focus group \#1} & \rotatebox{90}{RQ2 Maintainability} & \rotatebox{90}{Focus group \#2} & \rotatebox{90}{RQ1 Suitability} \\
            \midrule
1 & Software Engineer Lead   & \checkmark  & \checkmark    &      & \checkmark    & \checkmark       & \checkmark    & \checkmark   \\
2 & Software Engineer Lead   & \checkmark   & \checkmark  &      & \checkmark    & \checkmark       & \checkmark    & \checkmark   \\
3 & SaMD Quality Lead        & \checkmark   &             & \checkmark   & \checkmark    & \checkmark       & \checkmark    & \checkmark   \\
4 & Senior Software Engineer &      &             &      & \checkmark    & \checkmark       & \checkmark    & \checkmark   \\
5 & Project Manager          &      &             &      & \checkmark    & \checkmark       &       &              \\
6 & Test Manager             &      &             &      & \checkmark    & \checkmark       &       &              \\
\bottomrule
\end{tabular}}
\vspace{-0.5cm}
\end{table}

\emph{Initial case creation.}
Initially a generic SAC was created following the guidelines by Weinstock et al.~\cite{weinstock_arguing_nodate}. The context for this SAC was a web server running on a Raspberry Pi. This served as a basis for hands-on learning about SAC creation and a starting point for internal discussions. Based on feedback, another case was created with a similar context, this time using the CASCADE approach~\cite{mohamad2023cascade}. Creating these two cases helped  understanding the practical application of SAC and served as a reference in communicating to the participants what an SAC looks like and what it should contain.

\emph{Interviews.}
We conducted semi-structured interviews to collect qualitative information about the problem of interest. This approach also ensured that all interviews followed a similar pattern and thus generated data that was comparable between subjects.
In the first round, we interviewed two software engineering leads working on HOOP and the Software as a Medical Device (SaMD) Coordinator to get insights about the regulations and requirements placed on the documentation of the HOOP system and security demands from regulatory authorities. We also elicited important areas of the system as a foundation for the benchmark SAC.


The second round of interviews focused on the validation of findings in the documentation that were relevant to SAC. We interviewed the SaMD Coordinator as this is the role with the most knowledge about the specific requirements included in the standards relevant to the medical domain.

\emph{Benchmark case creation.}
To build the benchmark case, we used CASCADE to utilize its argumentation structure as well as the quality aspects that it integrates. Two authors collaborated with engineers from the case company with background knowledge of both the system under consideration and the medical domain to build the SAC in a top-down manner. This meant first identifying the relevant assets and the security goals, then looking at threats and risks, assigning treatments and requirements to the risks, and finally providing evidence for the requirements. The case was built in an iterative manner which allowed us to receive continuous feedback for the various components of the case from the domain experts. In the end, instead of aiming to create a \emph{complete} case, a partial SAC was created which contains enough information to assess the suitability of SAC in the medical domain as well as the suitability of CASCADE in agile environments. 
An excerpt of the SAC created for the benchmark case can be seen in Figure 6 in the supplemental material~\cite{supplemental-material}.

\emph{Documentation and regulation analysis.}
Regulations on HOOP from the medical domain require compliance with certain requirements for market approval. We assessed if SAC created with CASCADE fulfills these requirements and is suitable for the existing documentation. In our analysis, we investigated which parts of the documentation map to the building blocks of CASCADE (cf.~Section~\ref{sec:background:sac}) and which requirements in the regulations they fulfill.

\emph{Focus groups.}
We conducted two focus groups that were moderated by the first two authors. We used Morgan's method \cite{morgan} to allow for natural and spontaneous interactions and discussions among the focus groups' participants. The first focus group discussed a maintainability process for a SAC in the medical domain. The participants for the session were six experts with different roles and areas of responsibility as shown in Table~\ref{tbl:participants}, with varying degrees of expertise regarding the iterative workflow. We started with a short presentation that explained the topic as well as SAC and the approach used to create them. Following this, all participants filled in a questionnaire whose answers served as a basis for initial discussions. Afterward, the participants brainstormed suitable workflows and  their fits for key processes. The results of this focus group contributed to answering RQ2.

In the second focus group, we confirmed the results of the interviews, documentation analysis, and case creation and discussed the applicability of SACs in the medical domain and in agile development. Four participants with knowledge about the system of interest (cf.~Table~\ref{tbl:participants}) took part. We presented our preliminary results and discussed them with the participants. We also created use cases of SAC at the case company. These results contributed to RQ1 and RQ2.

\section{Suitability of SAC in the medical domain}
\label{sec:suitability}

This section discusses the use cases for SAC in the development of medical devices and the overlaps found in the documentation analysis. It therefore addresses \textbf{RQ1}.

\paragraph{Use cases for SAC in the medical domain}
The SaMD coordinator created 15 use cases for SAC in the medical domain. In addition, two more use cases were discovered during the second focus group for a total of 17. 
We identified use cases in a number of categories, most importantly \emph{compliance} and \emph{assessment} as well as \emph{planning} and \emph{monitoring}.
Examples of these usage cases can be seen in Table~\ref{tbl:use_cases}.
A full list is provided in Table 2 in the supplemental material~\cite{supplemental-material}. The numbers below refer to the IDs in the latter table.

\begin{table}
  \caption{Examples of identified use cases for SACs in the medical domain. The complete list is in the supplemental material~\cite{supplemental-material}.}
  \label{tbl:use_cases}
\setlength{\tabcolsep}{0.5em}
\resizebox{\columnwidth}{!}{\begin{tabular}{@{}p{.4cm}p{7.9cm}p{1.6cm}@{}}
\toprule
ID & Use Case Description & Category\\ 
\midrule
1 & 
  As a Device Regulatory Lead (DRL), I would use top-level SACs to prove to
  the regulatory agencies that the company has considered all relevant
  security aspects of the final product, and has enough evidence to
  claim that it has fulfilled them to meet the Intended Use claims of
  the medical device. 
  & Compliance \\ 
  \midrule


7 &
  As a product owner, I would use SACs to make an assessment of the
  quality of my product from a security perspective, and make a 
  roadmap for future security development. 
  & Monitoring, Planning \\
  \midrule



16 & 
  As a test manager I would use SACs in order to elicit what needs to be tested for a specific software artifact in order to facilitate traceability to user requirements.  
  & Planning \\ \midrule
17 & 
  As a SaMD Quality Manager, I would provide applicable evidence, e.g., test results to claims in the SAC case, in order to increase the quality and argumentative power of the SAC case, which in turn provides an increased ability to argue for the quality of the product. 
  & Assessment \\
  \bottomrule
\end{tabular}}
\vspace{-0.5cm}
\end{table}

In terms of compliance, several roles (e.g., the Device Regulator Lead, the QA Owner, and a member of the Risk and Cybersecurity team) can use SAC to show that the product is compliant with standards such as FDA guidance and ISO~14971 (1, 2, 3, 15). A precondition for this is a mapping between an SAC and the relevant standards, and we report below that SACs built with CASCADE cover a significant portion of the standards. 
In terms of assessment, project managers can determine whether security is at a stage where a product can be released (4), solution architects can use SACs to assess the quality of the product (11), similar to what product owners do (7). 

The quality aspect is also relevant for the SaMD Quality Manager: they can use the SACs to assess and iteratively increase the coverage of security tests and identify and validate new claims during development (17). This ties into the iterative way of working at the case company and illustrates the usefulness of SACs during the development phase. The participants in the focus group also mention that SACs can be reused between projects (12), providing new projects with a foundation of relevant security claims and evidence.

SACs can also be used for \emph{project planning and monitoring}. Several use cases address these issues from the point of view of different roles. On the one hand, this includes planning for resources to actually build the SAC (5), as well as identifying relevant next steps (10) and what concretely needs to be tested (16). SACs can also be used to track progress of suppliers (9) and if security requirements are fully covered (6). 

\paragraph{Overlap between SACs built with CASCADE and regulations for the medical domain}

CASCADE has a significant overlap with the most important regulatory standards and guidance documents in the medical domain. We investigated the SAMD pre- and postmarket guidance documents issued by the FDA, the pre- and post-market guidance documents issued by the MDCG, EMA, FDA, several ISO standards, and NIST 800-30. The overlap is shown in Table~\ref{tbl:standard-overlap}. More details are included in Figure 2 in the supplemental material~\cite{supplemental-material}.

\begin{table}[tp]
\caption{Important standards and guidance documents in the medical domain and their identified overlap with CASCADE.}
\label{tbl:standard-overlap}
\resizebox{\columnwidth}{!}{\begin{tabular}{@{}lllllllll@{}}
\toprule
         Reference          & Type  & \rotatebox{90}{Market}         & \rotatebox{90}{White hat} & \rotatebox{90}{Black hat} & \rotatebox{90}{Resolver}    & \rotatebox{90}{Generic} & \rotatebox{90}{Evidence} & \rotatebox{90}{Quality claims}       \\ 
\midrule
ISO 14971:2019   \cite{iso_14971}    & Standard       & Int.       & \checkmark & \checkmark & \checkmark & \checkmark & \checkmark  & \checkmark\\
ISO 62304:2006   \cite{iso_14971}    & Standard       & Int.       & \checkmark & \checkmark & \checkmark &            & \checkmark  &           \\
MDCG 2019-16    \cite{mdcg_guidance} & Guideline      & EU                  & \checkmark & \checkmark & \checkmark & \checkmark & \checkmark  &           \\
SaMD Premarket  \cite{premarket-fda} & Guideline      & US                  & \checkmark & \checkmark & \checkmark & \checkmark & \checkmark  & \checkmark\\
SaMD Postmarket \cite{postmarket-fda}& Guideline      & US                  & \checkmark & \checkmark & \checkmark &            & \checkmark  &           \\
NIST 800-30     \cite{nist800}       & Guideline      & US                  & \checkmark & \checkmark &            &            &             &           \\
EMA/226170/2021  \cite{gcp_guideline}& Guideline      & EU                  & \checkmark & \checkmark &            & \checkmark & \checkmark  &           \\
\bottomrule
\end{tabular}}
\vspace{-0.5cm}
\end{table}

As an illustration, we describe the overlap between CASCADE and ISO~14971~\cite{iso_14971}.
A central artifact of ISO~14971 is the ``risk management file''. This file contains documentation of intended usage and foreseeable misuse, identification of safety-related characteristics, and identification of hazards. 
Certain safety-related characteristics are expressed in CASCADE's White Hat Block, such as properties that need to be preserved (security goals) in order to prevent hazardous situations, namely security risks  that could have a safety impact. Threat scenarios that potentially result in a violation of the established properties that need to be preserved are described in CASCADE's Black Hat Block. Safety-related characteristics that have no connection to cybersecurity (such as short-circuits or battery damage) do not belong in a SAC and can be covered, e.g., with a fault tree analysis (FTA).

Risk control is performed indirectly through assigning risk treatments in CASCADE's Resolver Block and providing evidence that the risk is on an acceptable level. Providing traceability between risks and the ``risk management process'' is one of the core functionalities of CASCADE, as both aspects are explicitly combined in the SAC. The same reasoning applies to using the risk management file as a way to show compliance during an inspection. 

ISO~14971 also stresses the importance for completeness of risk management as ``an incomplete task can mean that an identified hazard is not controlled and harm can be the consequence.'' CASCADE addresses this by utilising quality claims that involve gathering evidence that all risks/hazards have been accounted for and arguing for the achieved completeness. It further calls for the identification of characteristics related to safety, as well as identification of hazards, which are tied to the identification of assets of the medical device. 

CASCADE's Generic Subcase can capture the ISO~14971 requirements on the competence of the personnel responsible for carrying out the risk management process and application of ISO~14971, and that they have the necessary skills, education, training for and experience of the applicable medical device, as well as of the technologies and the risk management techniques used during the risk management process.



\paragraph{SAC and Safety}
The norms referenced in the previous section explicitly require traceability between security issues and safety risks. First of all, they all require explicitly stating the safety-related risks and their mitigation. This is outside the scope of an SAC. The MDCG guidance, e.g., states that ``there is a need to consider the relationship between safety and security as they relate to risk [\ldots] safety may be compromised due to security issues which may have safety impacts''~\cite{mdcg_guidance}.

The FDA-issued postmarket guidance document states the need for assessing the severity of patient harm should a cybersecurity risk be exploited. It also references an approach outlined in ISO~14971, involving ``qualitative severity levels'', that can be used to conduct such an assessment. ISO~14971 also states that assessing and documenting the severity and probability of occurrence for risks with safety implications should be performed as part of risk estimation. It further states that manufacturers shall identify and document risks that may lead to hazardous situations (situations involving patient harm) during both intended use and foreseeable miss-use.

ISO~62304~\cite{iso_62304} states that manufacturers of medical device software are required to document a \emph{software safety class} for all software items (partial assets), that denotes the severity of the outcome should a hazard occur for that item. Assigning such classes is required in order to comply with the standard, as the class dictates which measures need to be be taken for a given software item to be deemed secure and safe. The need for these classes ties into the need for showing traceability in SACs between software security and the potential safety implications of security issues. 


It was proposed during the interviews that an approach for distinguishing safety-critical security claims in the case would be to incorporate the results from a ``risk assessment matrix''~\cite{risk_assessment_matrix}. This is an established system in the medical domain for the rating (on a scale from one to five) and flagging risks (using a color gradient from green to yellow and red) that have an inherent patient safety concern. These ratings and flags can be included in a SAC by assigning colors to identified safety-critical security claims in accordance to their calculated rating from the matrix. Adopting this would fulfill the need for SACs to provide a level of traceability for security risks with an inherent safety risk to patients.

Utilizing the same rating and color system used in other approaches in the medical domain for SAC has the benefit of not increasing the required labor for the case creation, as these ratings are being calculated anyway, as well as not increasing the complexity of understanding the case, as these ratings are already well established and used within the medical domain.


\section{Integrating the Maintenance of SACs Built with CASCADE into the Existing Workflow}

This section presents our results addressing \textbf{RQ2}.

\begin{figure*}[tb]
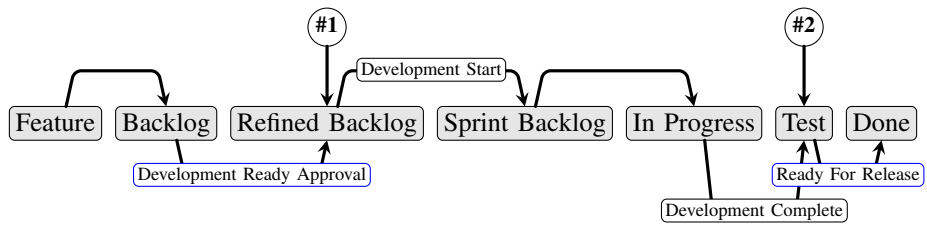

\centering
\depstyle{outer bubble}{fill=gray!20}
\begin{dependency}[edge style={very thick}]
    \begin{deptext}[column sep=.2cm, row sep=.1ex]
    Feature \&  Backlog \& Refined Backlog \& Sprint Backlog \& In Progress \& Test \& Done \\
    \end{deptext}

    \depedge[hide label]{1}{2}{Transition}
    \depedge[edge below, label style = {draw=blue, scale = 1 }]{2}{3}{Development Ready Approval}
    \depedge{3}{4}{Development Start}
    \depedge[hide label]{4}{5}{Transition}
    \depedge[edge height=6ex, edge below]{5}{6}{Development Complete}
    \depedge[edge below, label style = {draw=blue}]{6}{7}{Ready For Release}

    \deproot[edge unit distance=2ex, label style = {circle}]{3}{\large \textbf{\#1}}
    \deproot[edge unit distance=2ex, label style = {circle}]{6}{\large \textbf{\#2}}
    
    \wordgroup[outer bubble]{1}{1}{1}{feature}
    \wordgroup[outer bubble]{1}{2}{2}{backlog}
    \wordgroup[outer bubble]{1}{3}{3}{ref_backlog}
    \wordgroup[outer bubble]{1}{4}{4}{sprint_backlog}
    \wordgroup[outer bubble]{1}{5}{5}{in_progress}
    \wordgroup[outer bubble]{1}{6}{6}{test}
    \wordgroup[outer bubble]{1}{7}{7}{done}
\end{dependency}
    \caption{Overview of the HOOP platform workflow. Triage team approval gates are in \textcolor{blue}{blue}. The circles indicate suggestions for security-related extensions.}
    \label{fig:workflow}
    \vspace{-0.3cm}
\end{figure*}

The first focus group was mostly concerned with how to incorporate a maintainability process for SAC into the case company's current agile workflow. An overview of the feature flow can be seen in Figure~\ref{fig:workflow}.


The current workflow at the case company is built upon existing frameworks such as Scrum and phase-gate processes~\cite{pinto_project_2020}. Before a feature is deemed complete and ready for release, a comprehensive list of criteria needs to be fulfilled.

One of the major processes involved is ``triage approval''. Each member of the triage team, currently consisting of the PO, the ITPM, the Solution Architect, the Product Analyst, the ITQM, and the Squad Lead, needs to give individual approval at two points in time: 1) before a feature can be moved from the backlog to the refined backlog from where it can be selected for the sprint backlog; and 2) before a feature in the sprint backlog can be moved from ``Test'' to ``Done''. 
Each role in the triage team has their own set of criteria that need to be fulfilled before the role can approve the feature. 
The triage team does not currently include any security roles. 


The two places in which changes were suggested are shown as \textbf{\#1} and \textbf{\#2} in Figure~\ref{fig:workflow} and correspond to the points in time where the triage team gets involved. During the focus group there was a consensus that incorporating a maintainability process for SACs in the workflow requires several experts as none in the scrum teams would have all knowledge required to manage all blocks of CASCADE. One suggestion was to update the SAC as a part of triaging and adding a corresponding feature approval criterion, with a main focus on the White Hat block and asset identification.

First of all (\#1), when a feature is moved to the refined backlog as part of the design review, the product blueprint document usually needs to be extended. The blueprint contains a high level description of the architecture of the system, including the different assets the system is composed of. As features are added to the system, the blueprint needs to be extended with additional assets and these assets should also be considered in the SAC. Only if new assets have been added to both the blueprint and the SAC should the triage team pass the review and mark the feature as ready for development. This can be captured by a feature approval criterion and should be checked by the ``security architect'', a new role suggested by the focus group to be added to the triage team.


Secondly (\#2), after the feature has been marked as completed by the development team and is pending triage approval, the security architect should approve the suggested changes to the SAC. This means that the development team, supported by the security architect keeps the SAC up-to-date during development and review and update of the SAC could be added to the Definition of Done (DoD). This also means that the triage team only approves the feature if the SAC was updated.



\section{Discussion}
\label{sec:discussion}

As the results detailed in the previous sections show, security assurances cases built with CASCADE have a place in the agile development of medical devices. The feedback from the case company shows that they are suitable to continuously assess the system under construction in terms of its security and provides feedback that can be used in the ongoing development, e.g., in the context of test coverage or product quality. In the following, we will discuss our findings in the context of published literature and answer our research questions.






\paragraph{Applicability of SAC in the development of medical devices (RQ1)}

Our results indicate that SACs built with CASCADE are well suited for the medical domain. 
The use cases defined by the SaMD coordinator and by the focus group show that SACs are useful for several roles at the case company, both for external and internal needs. The interviews with the SaMD coordinator also show that SACs built with CASCADE provide the necessary documentation to comply with requirements from the relevant standards and guideline documents for cybersecurity in the medical domain. 

The documentation analysis confirms that all studied standards contain requirements that can be fulfilled with one or several parts of CASCADE. However, it also shows that there is a need to address safety-related security risks separately and that there is a need for traceability between security and patient safety concerns that stem from insufficient security measures.

In summary, the case study has shown that SACs can be applied in the medical domain and that CASCADE is a suitable tool for building them in the same context. 



We identified the need to distinguish safety-related security risks and purely security-based risks. 
This does not change the structure of the SAC created with CASCADE, but requires including the ratings and color from the ``risk assessment matrix''.
Such integration between security and safety has also been discussed in the literature, e.g., by Lisova et al.~\cite{lisova2018safety}. However, as Lisova et al. note, most existing approaches are limited as they are based on concrete regulations or on specific techniques with certain limitations. We argue that our approach of analysing the concrete needs of the case company and tailoring a technique-agnostic approach such as CASCADE to these needs is more viable. 
We furthermore subscribe to Ray and Rance's conjecture that increasing security has a positive impact on safety~\cite{ray_security_2015}, in particular when making the connections between the two areas explicit as we propose here.

\paragraph{Utilize and extend existing agile processes to maintain SAC (RQ2)}
Since the case company already develops safety-critical systems using an agile approach, incorporating security into the workflow as well is significantly easier than for companies that are not used to combining the rigor required by standards and regulations with the relative freedom of an agile approach. Concretely, this means that several of the aspects that Beznosov and Kruchten had identified as a ``mismatch'' between agile and secure software are already addressed in the current process~\cite{beznosov2004towards}. As an example, deriving requirements from specifications is a well-understood part of development at the case company and can just as easily be applied to security-critical systems as our results in Section~\ref{sec:suitability} show.

Many of the challenges listed by Oueslati et al.~\cite{oueslati2015literature} similarly do not apply to our case company since they already have structures in place for safety. Detailed documentation is already created and the organisation is aware of the impact new requirements and design changes can have on safety. These skills transfer directly to the area of security. Our proposed changes also address some of the challenges, e.g., the introduction of a security architect as a specialised role.

Likewise, Bartsch identifies mismatches between agile and secure system development in his interviews with practitioners~\cite{bartsch2011practitioners}. We also find that risk analysis needs to be a central part of secure system development and describe how using the risk assessment matrix already in place can support this activity. Bartsch also points to the need to continuously improve the security process. 
While process improvement is out of the scope for this paper, we have been discussing how the additional process steps proposed here can be further developed together with our contacts from the case company.

\section{Threats to validity}
This section discusses the threats to validity identified in this study as well as the methods used to minimize these threats. 

\subsection{Internal Validity}

The case company was not aware of the concept of SAC or of CASCADE before the study was conducted. Their interest in giving us access to their developers was to learn about both. However, the bulk of the knowledge of the participants about SAC and CASCADE came from the authors, potentially introducing a form of bias. However, most participants were familiar with other security techniques such as fault tree analysis which means that they had pre-existing knowledge of tools for risk management and risk analysis. The participants were also experienced with approaches for documenting safety and security properties and ensuring compliance with applicable standards, which means that they had a good understanding of the different needs security assurance approaches have.

\subsection{External Validity}
This section covers threats that relate to generalizability of the study and how well the results could be applied to a similar context outside of the study environment.

\paragraph{Generalizability}
As this case study has taken place at a single company building medical device software, the generalizability of the results might be limited. However, the documents and standards referenced are domain-wide, meaning that all companies building medical devices will have to consider and comply with these standards. Hence, we argue that the results from this case study can apply to other organizations in the domain that are driven by the same forces (in this case standards) \cite{generalize}.

\paragraph{Partial documentation analysis}
There are many standards and guidelines in the medical domain and each medical device or medical product is subject to a different subset. Based on our work with the SaMD compliance coordinator, we selected the subset that was applicable to the HOOP system. The standards mentioned in Section~\ref{sec:background:standards} do, however, cover many use cases. Therefore, while development teams should check which standards are applicable to their concrete product, our selection is representative for many cases.

The selected standards also have an innate cohesion by referencing and building upon each other, meaning that the requirements posed by the more specific standards not studied are a subset of the requirements in the studied ones or include requirements similar to the ones we have regarded.

\paragraph{Documentation and standard volatility}
Guidance documents from the FDA tend to be updated every two years, whereas ISO standards tend to be updated every five to six years. This means that the CASCADE approach might need to be revised regularly, depending on the amendments made to the document in question. During the period in which the study ran, no updates to the relevant guidances were published.

\paragraph{Result validation}
The results have been validated at one single company, and some were only validated by the SaMD compliance coordinator. This person has the deepest knowledge of applicable medical domain standards, and general knowledge of compliance requirements. However, the content of the standards and guideline documents were vetted by the authors before validation by the SaMD compliance coordinator, and confirmed in the second focus group.




\section{Conclusion}
\label{sec:conclusion}



In this paper, we evaluated how SACs built with CASCADE apply to the agile development of medical device software.
CASCADE is an approach to systematically create Security Assurance Cases and thus provide a thorough and systematic assessment of the security of a product and the way of working and qualification of an organisation.
To achieve this, we elicited input from domain experts in a case study that included interviews and focus groups, and analysed major regulations and standards from the medical domain. 
 
From this analysis, we derived that SACs are useful and CASCADE is applicable in the medical domain and in the case company. It provides several opportunities to assess the quality of the product and supports the established quality assurance mechanisms at the case company. A simple extension to highlight safety-related security risks using visual indicators according to the risk assessment matrix can provide traceability of safety risks.
In addition, incorporating SACs into the iterative, agile processes at the case company is feasible with certain additions. In particular, the Definition of Done has to be extended with statements about the completeness of the SAC and the new role of the security architect has to be added. This role has overall responsibility for the security documentation of the system and be a deciding factor in ensuring that the SAC is up-to-date before a feature enters development and before releasing the feature into production. It is also important that other roles involved in product development such as software developers support the maintenance of the SAC.

In future research, we intend to investigate the interplay between security and safety assurance. In particular, we plan to investigate the connection between SAC and established safety assurance methods such as FTA.


\bibliographystyle{splncs04}
\bibliography{references}

\end{document}


\title{Security Assurance Cases in the Medical Domain\\Supplemental Material}
%
%
\author{
Max Fransson\inst{2} \and
Adam Andersson\inst{1} \and
Mazen Mohamad\inst{1,2}\orcidID{0000-0002-3446-1265} \and
Jan-Philipp Steghöfer \inst{3}\orcidID{0000-0003-1694-0972}}
\authorrunning{Andersson, Fransson, et al.}
%
\institute{Chalmers University of Technology, Gothenburg, Sweden\\
\email{andersson@gmail.com}\and
Research Institutues of Sweden (RISE), Gothenburg, Sweden\\
\email{mazen.mohamad@ri.se}, \email{max.fransson@ri.se}  \and
XITASO GmbH, Augsburg, Germany\\
\email{jan-philipp.steghoefer@xitaso.com}}
\maketitle              

\section{Security Assurance Cases -- Example}

\begin{figure}[tb]
\centering
\includegraphics[width=0.65\textwidth]{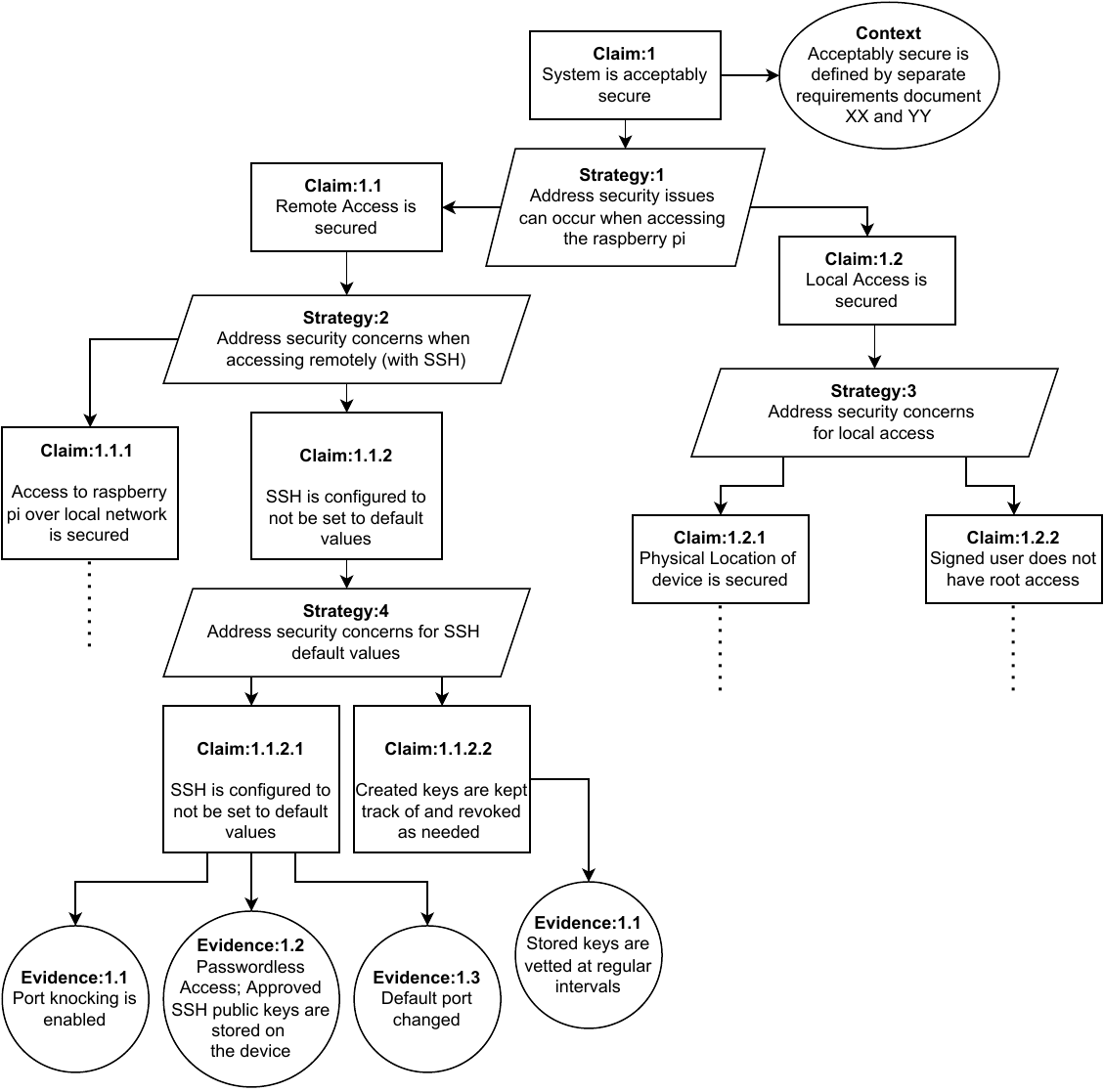}
\caption{A partial example SAC created for Raspberry Pi web server}
\label{fig:rasp_pi_sac}
\end{figure}

As an example for an SAC, a partial case for a Raspberry Pi web server is displayed in Figure~\ref{fig:rasp_pi_sac}, where the relation between claims, strategies and evidence can be seen. It also shows an instance of a context element which is used to help set the scope for the claim (where the scope can for example consist of product security requirements and standards that need to be complied with).

\section{Questionnaire for Maintainability Focus Group}
\label{sec:sup:maintainability-questionnaire}

\begin{itemize}
    \item What is your title at the case company?
    \item Have you heard of Security Assurance Cases before today?
    \item How familiar are you with agile work methodologies?
    \item What work methodology do you primarily use in your team?
    \item Are there any documents that you use to support your workflow? If so, which ones?
    \item Which standards or requirements or documents regarding security or safety do you use in your workflow?
    \item At which stage do you evaluate the security state of the system/product being developed?
    \item Do you have a suggestion on where in the process a security review would be suitable?
\end{itemize}

\section{Overview of Relevant Roles at the Case Company}

As HOOP is developed by a large multinational company, there are highly specific roles in place to fulfill the domain-specific operational needs and business needs, as well as the product development needs of the company. These roles are are outlined in Table \ref{tbl:sup:roles} below, with the role name, and a brief description of the responsibility of the role. 

\begin{longtable}{@{}P{4cm}p{8cm}@{}}
\caption{Overview of different relevant roles at the case company}
\label{tbl:sup:roles}\\
\toprule
Role & Responsibility\\ 
\midrule
\endfirsthead
\multicolumn{2}{c}%
{Table \thetable\ continued from previous page} \\
\toprule
Role & Responsibility\\ 
\midrule
\endhead
IT Project Manager &
  Agrees the SaMD life Cycle process with PO, DRP, ITQM \& SaMDQM, ITPMs and BPM. Responsible for the successful execution of all IT deliverables.\\
  \midrule
Product Owner (PO) &
  Ensures the product is developed to meet the business requirements. Owns, defines and prioritizes services/functionality that deliver business value. \\
  \midrule
SaMD Quality Manager (SaMDQM) Device Quality &
  Ensures activities that fall under SaMD procedures meet quality requirements and SaMD quality input into the project.\\ 
  \midrule
Business Quality Management (BQM) Clinical Quality &
  Reviews and approves non SaMD validation documentation in accordance with ITQMF. Ensures GCP compliance for validation activities. Provides GCP quality input into    the project. \\ 
  \midrule
Quality Technical Manager R\&D IT (SWQE) &
  Ensures SaMD quality processes are  implemented according to the quality strategy for the project/product \\ 
  \midrule
IT Quality Manager (IT QM) &
  Assures activities that fall under IT procedures meet IT quality requirements. Provides IT quality input into the project. \\
  \midrule
Device   Responsible Person (DRP) &
  Accountable for ensuring that development activities meet the medical device requirements. \\ 
  \midrule
Risk Management Lead (RML) &
  Responsible for establishment of the Risk Management File and maintenance throughout the development project. Reviews and approves relevant design controls deliverables. \\
  \midrule
Device Regulatory Lead (DRL) &
  Provides regulatory guidance and strategy for the development of the SaMD. \\ 
  \midrule
End-to-End Service Capability Manager (E2E CSM) &
  Responsible for developing service  and support models for deployment. \\
  \midrule
Operational Service Manager (OSM) &
  Service Management for the HOOP service. Plans and manages changes for the system or service. Monitors and manages the support processes. \\ 
  \midrule
Business   Analyst (BA) &
  Responsible for developing user requirements and ensuring requirements meet the business needs. \\
  \midrule
Solution Architect (SA) &
  Works with the business and BA  to understand functional, non-functional  and infrastructure requirements. Responsible for high-level design. \\
  \midrule
Test Lead (TL) &
  Responsible for creating test scripts and executing UAT and traceability to user requirements. Responsible for reviewing test scripts and executing UAT and  traceability to user requirements. \\
  \midrule
Test   Manager (TM) &
  Responsible and supporting test scripts creation and executing UAT and traceability to user requirements. Responsible and supporting and reviewing test scripts and executing UAT and traceability to user requirement. Responsible for generating final  test summary and traceability reports. \\
  \midrule
Configuration Manager (CM) &
  Handles the configuration management and updates the System index. \\
  \midrule
IT/SW Dev Lead (DEV) (Squad Lead) &
  Responsible for the Software Design, API Specifications, and Software Detail Design Plan. Responsible for code reviews and checklist compliance, unit and integration  tests execution and proper traceability. \\ 
  \midrule
Patient Safety Medical Device Lead (PSLMD) &
  Responsible for patient safety input. \\ 
  \midrule
IT Release Manager (ITRM) &
  Responsible for the successful execution of all IT deliverables during the release process. Manages release process for external/internal partners/clients
  \\
  \bottomrule
\end{longtable}

\section{Use Cases of CASCADE in the Medical Domain}

The SaMD coordinator created a total of 15 use cases for SAC in the medical domain seen in Table~\ref{tbl:sup:use_cases}. Two more use cases (numbers 16 and 17 in Table~\ref{tbl:sup:use_cases}) were elicited during the second focus group.

\begin{longtable}{@{}p{.6cm}p{9.4cm}p{2cm}@{}}
\caption{Overview of identified use cases for SACs}
\label{tbl:sup:use_cases}\\
\toprule
ID & Use Case Description & Category\\ 
\midrule
\endfirsthead
\multicolumn{2}{c}%
{Table \thetable\ continued from previous page} \\
\toprule
ID & Use Case Description \\
\midrule
\endhead
1 & 
  As a Device Regulatory Lead (DRL), I would use top-level SACs to prove to
  the regulatory agencies that the company has considered all relevant
  security aspects of the final product, and has enough evidence to
  claim that it has fulfilled them to meet the Intended Use claims of
  the medical device. 
  & Compliance \\ 
  \midrule
2 & 
  As a member of the Risk and Cybersecurity teams, DRP \& RML, I would
  use detailed SAC to prove compliance to the company, and regulatory teams that
  the project has complied to the company's Risk \& Cybersecurity standard operationg procedure, ISO 14971 standard, FDA Guidance in addressing patient  safety and cybersecurity  concerns and show them evidence of my claim of compliance. 
  & Compliance \\
  \midrule
3 & 
  As a member of the compliance team, SaMDQM, BQM, SWQE, ITQ
  I would use detailed SACs to review and ensure compliance to the company's Risk
  \& Cybersecurity standard operating procedures, ISO 14971 standard, FDA Guidance in addressing patient safety and cybersecurity concerns and document the
  effectiveness of the QA review.  
  & Compliance \\
  \midrule
4 & 
  As a project manager or RM, I would use SACs to make sure that a project is
  ready from a security point of view to be closed and shipped to production.  
  & Assessment \\
  \midrule
5 & 
  As a project manager or RM, I would include SACs in my project plan. I
  would make sure the project has the needed resources and time for 
  creating the case (argumentation, evidence collection, etc).
  & Planning \\ 
  \midrule
6 & 
  As a project manager I would use SACs to monitor the progress of my 
  project when it comes to fulfillment of security requirements.  
  & Monitoring \\ 
  \midrule
7 &
  As a product owner, I would use SACs to make an assessment of the
  quality of my product from a security perspective, and make a 
  roadmap for future security development. 
  & Monitoring, Planning \\
  \midrule
8 & 
  As a product owner, responsible for my project’s handling threats 
  and vulnerabilities, I would use SACs to evaluate the effect of new 
  threats and vulnerabilities, and evaluate whether a change is needed
  to the product or product lines.
  & Planning \\
  \midrule
9 & 
  As a member of the supplier assessment team, I would include SACs as 
  a part of the contracts made with suppliers, in order to have evidence
  of the fulfillment of security requirements at delivery time, and to
  track progress during the supplier’s development time.
  & Monitoring \\
  \midrule
10 & 
  As an cybersecurity Subject Matter Expert, SME on a project team,
  I would use detailed and visual SACs to communicate with the risk 
  owner, and decide how to update the product security in the right way
  (to know what to do). 
  & Planning \\
  \midrule
11 & 
  As a system leader or solution architect on a project, I would 
  use SACs to make an assessment of the quality of my system from a 
  security perspective, and make a roadmap for future security 
  development. 
  & Assessment \\ 
  \midrule
12 & 
  As a software developer responsible for implementing cybersecurity 
  controls on my project, I would use SACs from previous similar projects
  as a guideline for secure development practices. 
  & Reuse \\ \midrule
13 & 
  As a corporate QA owner, I would use SACs during a EU or FDA inspection
  if a regulatory issue is raised against the company for security 
  related issues. I would use the SAC to prove that sufficient preventive 
  actions were taken. 
  & Compliance \\ \midrule
14 & 
  As a member of the corporate communication team, I would use SACs as a 
  reference to answer security related questions.
  & Assessment \\ \midrule
15 & 
  As a member of the Patient Safety team, I would use detailed SACs to 
  prove to the company compliance and effectiveness of addressing any patient safety 
  concerns with respect to cybersecurity threats that have the potential 
  of any patient safety concerns.
  & Compliance \\ \midrule
16 & 
  As a test manager I would use SACs in order to elicit what needs to be tested for a specific software artifact in order to facilitate traceability to user requirements.  
  & Planning \\ \midrule
17 & 
  As a SaMD Quality Manager, I would provide applicable evidence (test cases and test suite results) to claims in the SAC case, in order to increase the quality and argumentative power of the SAC case, which in turn provides an increased ability to argue for the quality of the product. 
  & Assessment \\
  \bottomrule
\end{longtable}

\section{Overlap of CASCADE with Existing Regulation}

\begin{figure}[tb]
\centering
\includegraphics[width=\textwidth]{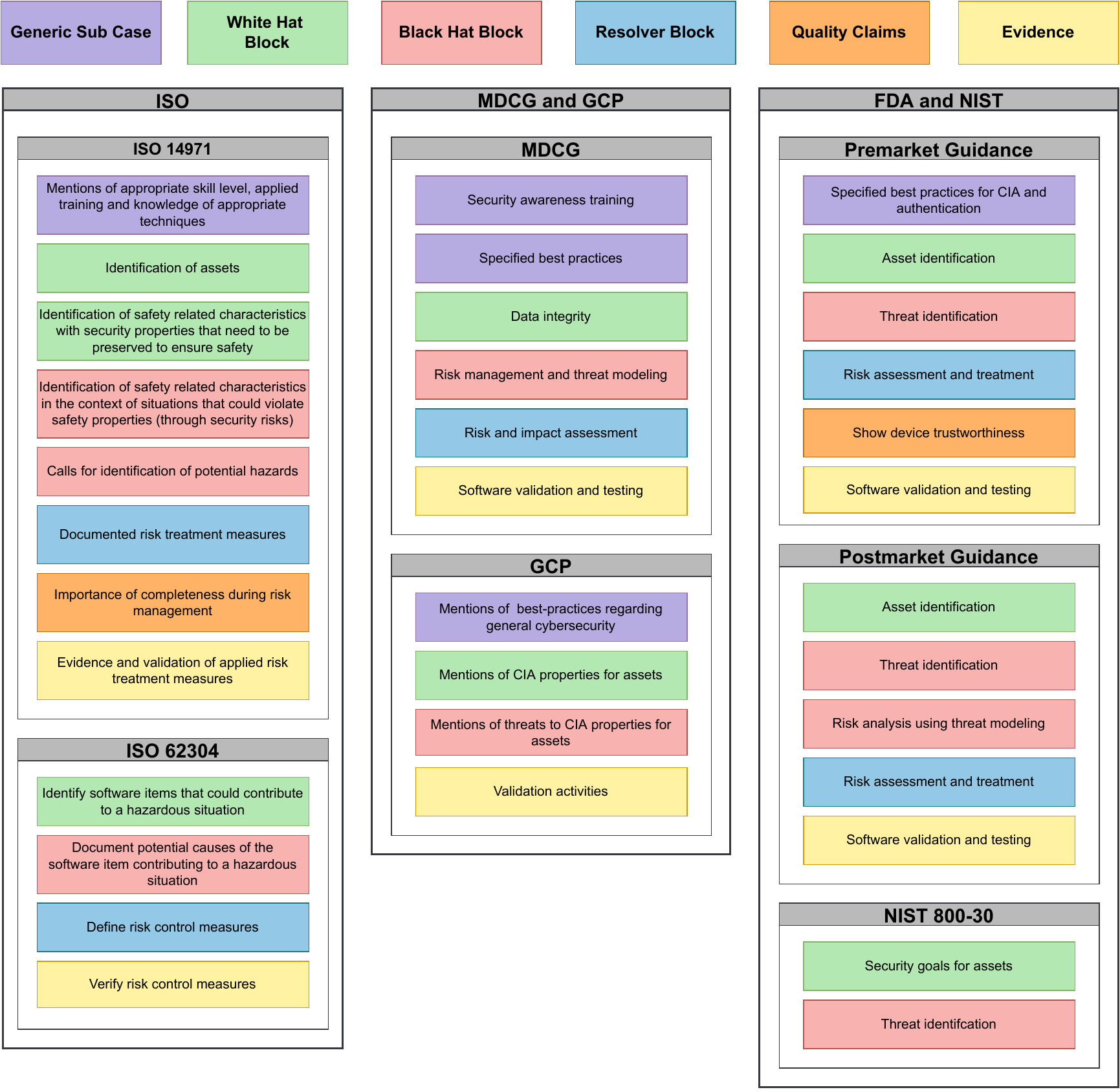}
\caption{High level overview of findings in the documentation and their overlap with the blocks of CASCADE. Each concrete part of CASCADE has a specific color, and the corresponding finding in each document that relates to that concrete part has the same color.}
\label{fig:sup:doc_overview}
\end{figure}

\subsubsection{Post market}
In the postmarket management document (PMD) \cite{postmarket-fda} there are several requirements stated that directly tie into different parts of CASCADE. Early on, it states that manufacturers are required to have a ''cybersecurity vulnerability and management approach'' \cite{postmarket-fda} in place. They then proceed to outline the concrete parts required for such an approach. These are as follows: 

\vspace{0.05cm}

\begin{enumerate}
    \item Identification of assets, threats, and vulnerabilities; 
    \item Assessment of the impact of threats and vulnerabilities on device functionality and end users/patients; 
    \item Assessment of the likelihood of a threat and of a vulnerability being exploited; 
    \item Determination of risk levels and suitable mitigation strategies; 
    \item Assessment of residual risk and risk acceptance criteria. 
    \label{hejsan}
\end{enumerate} 

\vspace{0.05cm}

Looking at the first item there is a clear use case for the white hat block and the black hat block, with a special emphasis on the first level of both, these being, for the white hat block \textit{“asset identification and decomposition”}, and for the black hat block, \textit{“threat scenarios”}. These are, as the names imply, used for identifying and breaking down the assets that compose the system, and stating what the potential threats to these assets are.

When looking at the second, third, fourth and fifth item there is a connection to the resolver block, which handles risk assessment, and what the treatment for a specified risk should be (mitigation, transfer etc.).

The PMD also states a recommendation for manufacturers to perform cybersecurity risk analyses, and as a part of this analysis, the inclusion of threat modeling is proposed. As explained by Mohamad et al. \cite{mohamad_asset-driven_2021}, a \textit{Threat Assessment and Remediation Analysis} (TARA) \cite{TARA} performed using the threat model STRIDE \cite{stride_owasp}, was used to create the black hat block in a case study at Volvo, which indicates that a TARA, like the one proposed by this document, could be expressed in terms of a black hat block in CASCADE. 

It also urges manufacturers to perform software validation, taking the form of software testing (such as unit tests, integration tests etc.), in conjunction with the previously mentioned risk analysis. It then proceeds to elaborate on the main purpose of the software validation, this being the assurance that the remediation applied to identified risks was successful. Not only would CASCADE provide a logical approach for displaying these tests and their results through evidence, but it would also allow for the designer of the case to specify the connection between the risks and their specific testing using risk treatment and concrete requirements as an intermediary step. 

\subsubsection{Premarket}

The premarket management document (PreMD) \cite{premarket-fda}, contains the same, identical, bullet list (bullet list \ref{hejsan}) as the one presented in the PMD and so the same connections that were made between the PMD and CASCADE related to that list can be made for this document as well.

However, something that is stressed in the PreMD that is not brought up in the PMD, is the need for devices to be trustworthy, stating that ''Manufacturers should design trustworthy devices and provide documentation to demonstrate the trustworthiness of their devices in premarket review'' \cite{premarket-fda}. CASCADE has the means to provide this documentation, but a prerequisite to this is that the system already needs to be adequately tested and verified, since the role of SAC is to provide documentation that demonstrates what security measures have already been taken. Having performed rigorous testing and risk proofing of a system does not provide trustworthiness itself, unless it can be properly portrayed though proper documentation, validation and argumentation. This is where CASCADE has the ability to create confidence in cybersecurity contributing towards trustworthiness trough the case itself, as this conveys all known and relevant (as scoped in terms of ''acceptably secure'') cybersecurity measures taken and through the use of “case quality assurance”. ''Case quality assurance'' is an element to CASCADE that tries to verify that breakdowns made in the case are exhaustive/complete, meaning that all assets, risks, mitigation strategies etc. have been identified and accounted for. It is also used to show that claims with evidence assigned uphold a certain amount of quality. These two aspects serve to create more trustworthiness in the documentation, which in turn helps to provide trustworthiness for the device.

Just as with the PMD, there are several mentions of software validation and why this is necessary, stating reasons such as reasonable assurance of the safety and effectiveness for the system/product in question. However, the PreMD goes a step further than the PMD by outlining specific design implementations that they recommend (for the submission to be approved by FDA). 

The kind of specific implementations include:
\begin{figure}[!ht]

\begin{center}
\begin{large}

``Limit access to devices through the authentication of users'' \cite{premarket-fda}\\

\vspace{15pt}

``Verify the integrity of all incoming data'' \cite{premarket-fda} 
\end{large}
\centering
\caption{Requirements taken from the PMD document by the FDA~\cite{premarket-fda}}
\label{fig:sup:requirements-pmd}
\end{center}
\end{figure}

\vspace{3pt}

Given a concrete list of required implementations (with a base in cybersecurity, like in Figure~\ref{fig:sup:requirements-pmd}), any potential SAC approach could prove beneficial for demonstrating compliance with these. The list of implementations can first be used together with the context element, to help set the scope of the case, and further aid in defining what the reoccurring ''acceptably secure'' means for the case. If this list has been kept in mind during the product creation, then the case can be used to show that these implementations truly have been taken into consideration for all relevant assets.

Finally the PreMD outlines a list of best practice activities, such as password handling and user authentication. Adherence to best practice elements like these that are relevant throughout the company and incorporate several products can be documented in the generic subcase part of CASCADE.

\subsubsection{NIST 800-30}

The NIST 800-30 guidance document outlines a risk management process that includes identifying security goals for assets, identifying vulnerabilities, and gives examples of concrete attack vectors (such as phishing attacks, DDoS attacks), in conjunction with suggestions on how the severity of attack vectors can be measured. The identification of assets and security goals align with the purpose of the white hat block in CASCADE, and the identification of vulnerabilities and attack vectors has an overlap with the black hat block in CASCADE.



\subsubsection{Medical Device Coordination Group 2019-16}
The MDCG guidance document \cite{mdcg_guidance} contains similar connections to CASCADE as the FDA issued guidance documents. It explicitly points out the usage of risk management system and threat modeling and security verification and validation through testing. There are also a requirement taken from Annex 1 in the same document regarding information security stating that devices incorporating software should be developed using state of the art risk management and verification. Along with this they provide a definition from ENISA for the information security domain ``Protection against the threat of theft, deletion or alteration of stored or transmitted data within a cyber system''. This definition ties in to all parts of CIA and through extension the security goals in the white hat block.

However, the MDCG also includes a statement that urges healthcare providers to learn and adhere to best practices when it comes to general cybersecurity measures. A list from the MDCG document~\cite{mdcg_guidance} of what is meant by best practices is specified in the list below:   

\begin{enumerate}
    \item Good physical security to prevent unauthorised physical access to medical device or network access points.
    \item Access control measures (e.g. role based) to ensure only authenticated and authorised personnel are allowed access to network elements, stored information, services and applications.
    \item Network access controls, such as segmentation, to limit medical device communication.
    \item General patch management practices that ensure timely security patch updates.
    \item Malware protection to prevent unauthorised code execution;
    \item Security awareness training.
    \item Auditability that supports non-repudiation, i.e. the ability to reliability determine who made what changes to the system and when to assist with forensics.
\end{enumerate}

Viewing these from the perspective of CASCADE, a strong connection can be made between the implementation of these, and the block of CASCADE known as the ''generic sub case''. As the main purpose of the generic sub case is to abstract and document general practices and standard operating procedures that are prevalent at the entity in question (such as a pharmaceutical company), that in turn lead to an improved security at the entity. The incorporation and conformity of the above mentioned items would be displayed in the generic sub case. This is due to all of them relating to general non-product/system specific practices (or practices that span over for example a family of products). 

In the MDCG guidance document, there is also a section dedicated to how documentation should be handled. In this section it specifically states that documentation should conform with the requirements stated in ``Medical devices regulations, Annex I''. Showing that conformity has been achieved given specific regulations is one of the key features of a SAC approach, and as explained earlier having concrete requirements are crucial for scoping the case (which is immensely important when the main focus is demonstrating compliance or conformance to specific requirements, as SACs in general have a tendency to grow very large when the scope is large). Looking closer at Annex I \cite{mdcg_guidance}, there are requirements that regard risk and impact assessment, security awareness training and data integrity that would be part of the resolver block, generic sub case and white hat block in that order.

\begin{figure}[h]
\centering
\includegraphics[width=0.5\textwidth]{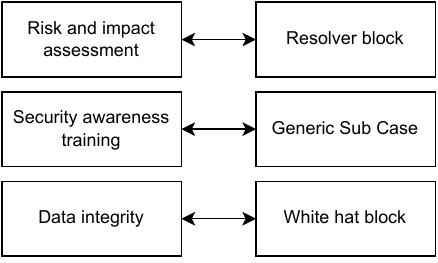}
\caption{Mapping between important areas introduced in MDCG Annex I and CASCADE blocks}
\label{fig:sup:mapping_mdcg_cascade}
\end{figure}


\subsubsection{Good Clinical Practice (GCP)}

In response to an enquiry made to the European Medicines Agency, regarding the cybersecurity requirements imposed by the GCP quality standard, the response outlined important sections of the ``Guideline on computerised systems and electronic data in clinical trials''. 

It mentions an array of best-practices for a good cybsersecurity posture, including the use of data backups, network firewalls, anti-virus software and authentication policies. At one point the sentence ``There should be documented training on the importance of security'' is used. Showing conformance with best practice requirements like these can be achieved through the generic sub case. The best practices present in the document are motivated through threat scenarios/attacks that could occur if the implementation of the best practices is not successful.

It further stresses the importance of data integrity, as any unauthorized or uncontrolled changes to the data can jeopardize the results of the clinical trial, as well as impact the privacy and integrity of the participants in the clinical trial. This property is something that could be shown through the white hat block (through security goals, i.e the integrity is not compromised).

\subsubsection{ISO 14971}
A central piece to the ISO 14971 \cite{iso_14971} is an artefact called the ``risk management file''. The way this file is specified ties into several of CASCADEs usage areas. The specification consists of a risk analysis defined as documentation of intended usage and foreseeable misuse, identification of safety related characteristics and identification of hazards. 

Certain safety related characteristics can then be expressed in the white hat block, such as properties that need to be preserved (security goals) in order to prevent hazardous situations, namely security risks with properties that could have a safety impact. There are also safety related characteristics that can be expressed in the black hat block, but then of the kind that relates to the concrete situations that could potentially result in a violation of the established properties that need to be preserved (threat scenarios). However, there are some safety related characteristics that have no connection to cybersecurity (such as short-circuits or battery damage) that do not belong in a SAC.

Risk control is indirectly performed through assigning risk treatments in the resolver block and then providing evidence, that ensures that the risk has reached an acceptable level. In regards to the requirement of providing traceability between risks and the ``risk management process'', it is one of the core functionalities of CASCADE and any other SAC approach, as they are explicitly tied together in the assurance case. The same reasoning goes for using the risk management file as a way to show compliance during an inspection. 

The document also stresses the importance for completeness when doing risk management as ''An incomplete task can mean that an identified hazard is not controlled and harm can be the consequence.'' CASCADE tries to control for this by utilising quality claims that involves gathering evidence that all risks/hazards have been accounted for, and argues for the achieved completeness. It further calls for the identification of characteristics related to safety, as well as identification of hazards, which are tied to the identification of assets that the medical device consists of. 

There are also mentions the requirement of competence of the personnel responsible for carrying out the risk management process and application of ISO 14971, and that they have the necessary skills, education, training and experience of the applicable medical device, as well of the technologies and the risk management techniques used during the risk management process. These properties tie in to the ``Generic Sub Case'' of CASCADE, and that there are skills and practices in place that carry over between different (separate) applications of SAC creation using CASCADE.

A noteworthy statement that ISO 14971 makes is that the standard can be used to assess all types of risks that are related to medical devices, not only cybersecurity related ones. As CASCADE is a SAC approach, only cybersecurity based risks are to be recorded in the case, meaning that purely safety based risks need to be handled with another process (such as FTA and SAAC).

\subsubsection{ISO 62304}

As ISO 62304 requires that a risk management process is applied in accordance with the specifications in ISO 14971 ''The \textbf{MANUFACTURER} shall apply a \textbf{RISK MANAGEMENT PROCESS} complying with ISO 14971.'', the same connections to CASCADE that was made for that standard in regards to the risk management process can be made for ISO 62304 as well.

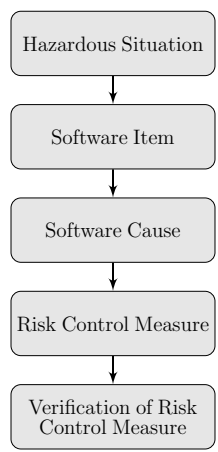
\begin{figure}[h]
\centering
    
\begin{tikzpicture}[node distance = 0.4cm, auto]
    \node [block, scale=0.7] (init) {\large Hazardous Situation};
    \node [block, scale=0.7, below=of init] (identify) {\large Software Item};
    \node [block, scale=0.7, below=of identify] (mutate) {\large Software Cause};
    \node [block, scale=0.7, below=of mutate] (transfer) {\large Risk Control Measure};
    \node [block, scale=0.7, below=of transfer] (evaluate) {\large Verification of Risk Control Measure};
    \path [line, thick] (init) -- (identify);
    \path [line, thick] (identify) -- (mutate);
    \path [line, thick] (mutate) -- (transfer);
    \path [line, thick] (transfer) -- (evaluate);

\end{tikzpicture}
\caption{Outline of risk management process flow in ISO 62304 \cite{iso_62304}}
\label{fig:sup:62304_flow}
\end{figure}

The ISO 62304 standard also requires that documentation providing traceability be created, the sought after traceability is described as: hazardous situation -> software item -> software cause -> risk control measure -> verification of measure (as shown in Figure~\ref{fig:sup:62304_flow}). This way of showing traceability is almost identical to the flow in CASCADE going from an asset (software item) in the white hat block, through the other blocks, to finally reach a level where evidence is provided.

\section{Security-Assurance Case for the HOOP System}

\begin{figure}[tb]
\centering
\includegraphics[width=\columnwidth]{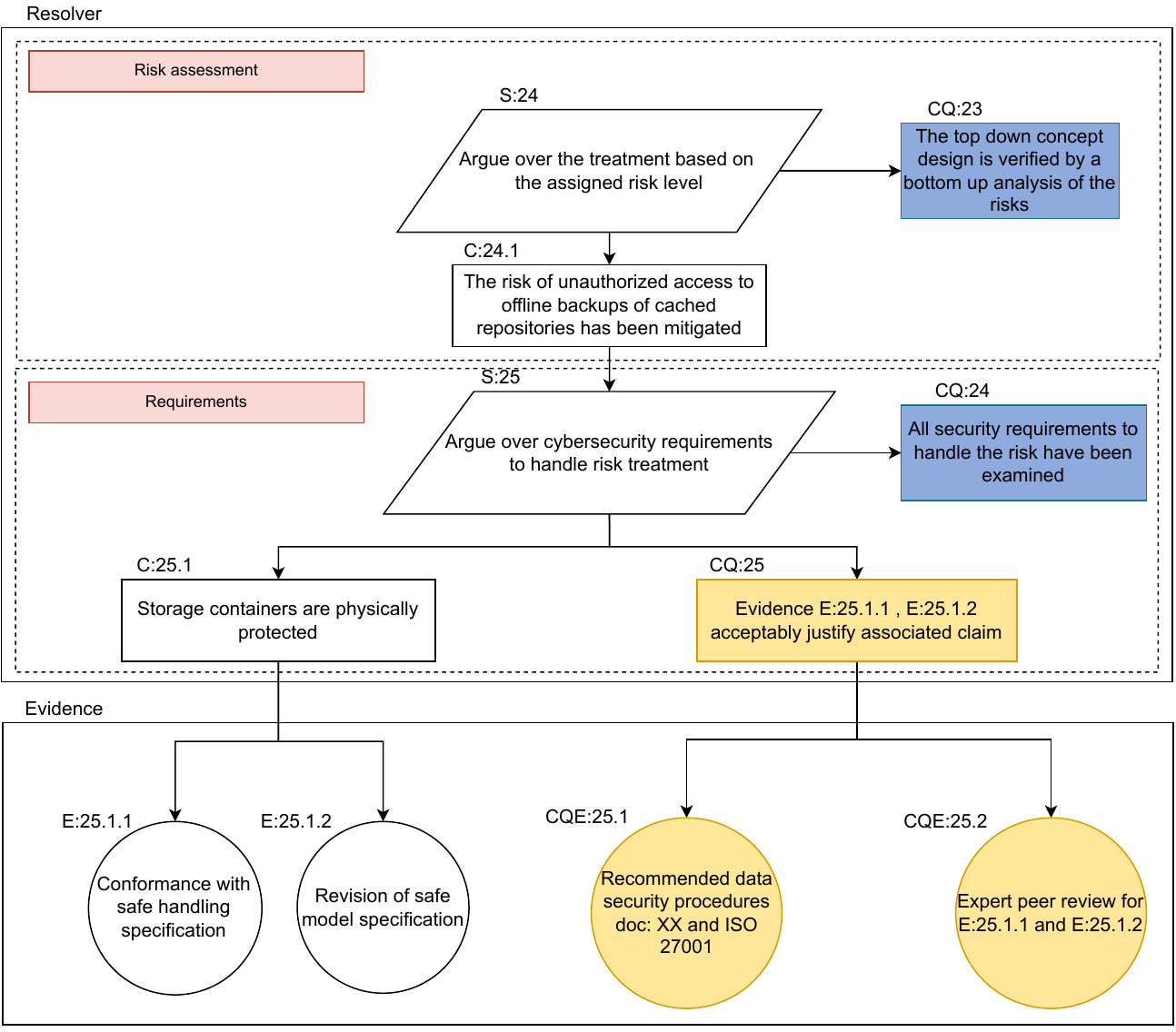}
\caption{Excerpt of the SAC created for the HOOP system.}
\label{fig:sup:res_sacExample}
\end{figure}

Figure~\ref{fig:sup:res_sacExample} shows an excerpt from the benchmark case created for the HOOP system. The arguments and evidence shown in the figure are taken from the resolver and evidence CASCADE blocks. The example in the figure can help test managers in use case (16) to assess if there is sufficient evidence to justify the claims in the requirements level of the resolver block, e.g., by examining the case quality evidence claims CQE:25.1 and CQE:25.2. 

\bibliographystyle{splncs04}
\bibliography{references}